\documentclass[12pt]{spieman}  
\usepackage{amsmath,amsfonts,amssymb}
\usepackage{graphicx}
\usepackage{setspace}
\usepackage{tocloft}
\usepackage{lineno}
\usepackage{gensymb}
\usepackage{hyperref}
\usepackage{xspace}
\usepackage[normalem]{ulem}
\usepackage[dvipsnames]{xcolor}
\newcommand{\addtext}[1]{\textcolor{Emerald}{#1}}

\newcommand{\rsc}{\ensuremath{R_{\rm aSil/aC}}\xspace}
\newcommand{\umin}{\ensuremath{U_{\rm min}}\xspace}
\newcommand{\omegastar}{$\Omega_*$\xspace}
\newcommand{\yasil}{$Y_{\rm aSil}$\xspace}
\newcommand{\ylcm}{$Y_{\rm lCM}$\xspace}
\newcommand{\yscm}{$Y_{\rm sCM}$\xspace}
\newcommand{\dustem}{{\sc dustem}\xspace}
\newcommand{\themis}{{\sc themis}\xspace}

\title{The radial variation of the silicate-to-carbon ratio in M31 probed by PRIMA}

\author[a]{J\'er\'emy Chastenet}
\author[a]{Ilse De Looze}
\author[a]{Maarten Baes}
\author[b]{Simone Bianchi}
\author[c]{Viviana Casasola}
\author[d]{Laure Ciesla}
\author[e]{Stephen Eales}
\author[f]{Jacopo Fritz}
\author[g]{Fr\'ed\'eric Galliano}
\author[g]{Suzanne C. Madden}
\author[h]{Angelos Nersesian}
\author[i,j]{Monica Rela\~no}
\author[e]{Matthew W. L. Smith}
\author[a]{Stefan van der Giessen}
\author[h]{Emmanuel Xilouris}

\affil[a]{Sterrenkundig Observatorium, Universiteit Gent, Krijgslaan 281-S9, 9000 Gent, Belgium}
\affil[b]{INAF – Osservatorio Astrofisico di Arcetri, Largo E. Fermi 5, 50125 Firenze, Italy}
\affil[c]{INAF - Istituto di Radioastronomia, Via Piero Gobetti 101, 40129 Bologna, Italy}
\affil[d]{Aix Marseille Univ, CNRS, CNES, LAM, Marseille, France}
\affil[e]{School of Physics \& Astronomy, Cardiff University, The Parade, Cardiff CF24 3AA, UK}
\affil[f]{Instituto de Radioastronomía y Astrofísica, Universidad Nacional Autónoma de México, Morelia, Michoacán 58089, Mexico}
\affil[g]{Universit\'e Paris-Saclay, Université Paris Cité, CEA, CNRS, AIM, 91191, Gif-sur-Yvette, France}
\affil[h]{STAR Institute, Quartier Agora – All\'ee du six Ao\^{u}t, 19c B-4000 Li\`ege, Belgium}
\affil[i]{Dept. Fisica Teorica y del Cosmos, E-18071 Granada, Spain}
\affil[j]{Instituto Universitario Carlos I de Fisica Teorica y Computacional, Universidad de Granada, E-18071 Granada, Spain}
\affil[h]{Institute for Astronomy, Astrophysics, Space Applications \& Remote Sensing, National Observatory of Athens, P. Penteli, 15236 Athens, Greece}

\cftpagenumbersoff{figure}
\cftpagenumbersoff{table} 
\begin{document} 
\maketitle

\begin{abstract}
The properties of interstellar dust grains are being scrutinized more than ever before, with the advent of large facilities. Infrared emission from dust grains is a powerful asset than can help constrain their physical and chemical properties. Among these, the relative ratio of carbon-rich to silicate-rich grains remains one that has not yet been investigated thoroughly, due to the lack of dedicated instruments and modeling limitations. 
In this paper, we quantify the modeling degeneracies inherent to constraining the far-infrared (far-IR) slope of the dust emission spectral energy distribution. Used as a proxy for the silicate-to-carbon ratio, we find that recovering the far-IR slope is affected by the estimate of the local radiation field, and the input abundances of different grain species.
We show that PRIMA's hyperspectral imaging will lead to better constrained local radiation fields which will aid---together with PRIMA's polarization capabilities---to better constrain the silicate-to-carbon ratio in M31, and how it spatially varies within the galaxy.
\end{abstract}


{\noindent \footnotesize\textbf{*}J\'er\'emy Chastenet,  \linkable{jeremy.chastenet@ugent.be}; Ilse De Looze,  \linkable{ilse.delooze@ugent.be} }

\begin{spacing}{2}   

\section{Introduction}
\label{sect:intro}  
Dust in the interstellar medium (ISM) of galaxies is a key component in regulating the energy distribution in its local environment. Dust grains absorb ultraviolet (UV) and optical light from young stellar populations, which increases their temperature, and re-emit that light in lower energy wavelengths, in the infrared (IR), as they cool down. This leads to (i) a significant decrease in the photons' energy, making dust grains act as a coolant, and (ii) a nuisance when it comes to estimating stellar-related quantities, such as stellar mass or star formation rate\cite{DeLooze2014, Nersesian2020, Nersesian2020M51, Verstocken2020, Viaene2020}, as a major part of that light can be extinguished and reddened by dust grains\cite{Viaene2016, Bianchi2018}.
Dust grains also participate in several physicochemical processes in the ISM. They act as a catalyst for the formation of H$_2$ \cite{LeBourlot2012, Bron2014}, leading to the formation of molecular clouds necessary for star formation, and the smallest grains account for most of the photoelectric effect heating the gas \cite{Wolfire95} (and competing with the cooling effect of the larger grains). 
For these reasons, understanding the life-cycle, formation, and destruction of dust grains is critical so that we can not only correct for their effect on other galactic properties, but also better understand galactic evolution as a whole.

To do so, we often measure dust properties comparing observations with models that are meant to encompass physical properties of interstellar dust: size distribution, composition, abundance, etc. This approach can be applied either to extinction measurements, the combined effects of absorption and scattering as a function of wavelength, or emission measurements, from mid-infrared (mid-IR) to sub-millimeter (sub-mm) wavelengths, on which we focus in this paper.
Emission models can be split between rather simplistic analytical models (e.g., modified blackbody, reproducing well the emission from large grains, in thermal equilibrium\cite{Gordon2014, MeisnerFinkbeiner2015}) and physically-motivated, either observationally- or laboratory-based models\cite{Zubko2004, DL2007, Jones2013, THEMIS}. This last class is used when the spectral coverage samples the emission spectral energy distribution (SED) also in mid-IR wavelengths, in addition to the far-infrared (far-IR)-emitting blackbody regime.

A substantial amount of work in the past few decades has led to some solid knowledge of interstellar dust. It is made up of grains showing a distribution of sizes, from a few nm to about $1~\mu$m in size for the largest. 
It shows features associated to the presence of amorphous silicate-rich material, with metallic nano-inclusions to account for depletions, as well as features arising from vibrational bonds of aromatic-rich material, and a population of amorphous carbon-rich grains. 
These grains are also likely not spherical but elongated or irregular in shape, as suggested by the detection of polarized light, caused by preferential alignment of non-spherical grains along the magnetic fields. These are the main properties we believe useful to mention for this paper, that virtually all models agree on.
However, the details of all these parameters may differ from observation to observation. Spatially integrated and resolved studies of nearby galaxies have shown a wide range of variations between and within galaxies, proving that interstellar dust is an evolving component\cite{MunozMateos2009, Boselli2010, Ciesla2014, RemyRuyer2014, Hunt2015, Davies2017, Aniano2020, Galliano2021, Abdurrouf2022, Casasola2022, Dale2023, Chastenet2025}.

Using physical models, our goal is to investigate the possibility to recover a silicate-to-carbon ratio for the larger grain populations. 
Being able to disentangle silicate and carbon dust content would inform us on their relative spatial distribution, which could be linked to formation and destruction processes. For example, silicate-rich material is believed to form in O-rich Asymptotic Giant Branch (AGB) stars, while carbon-rich grain would be formed in C-rich AGB stars\cite{Gail2009, Goldman2022}. There is also theoretical work that suggest that carbon dust and silicate dust are not similarly sensitive to destruction processes in supernova environments\cite{Bocchio2014, Slavin2015, Hu2019}, although the literature is not unanimous. Understanding the differential evolution of carbon- vs silicate-rich dust grains would give us great insights on the grain life-cycle in the ISM.
This approach is heavily motivated by the sometimes peculiar shape of far-IR to sub-mm SEDs of some galaxies, which is sensitive to the silicate-to-carbon ratio. For example, the Small Magellanic Cloud exhibits, in certain regions, a shallower slope in this regime compared to other galaxies\cite{Bot2010}. This may also be linked to the ``sub-mm excess'' observed in the residuals of infrared modeling (when comparing data to models)\cite{Galliano2011, Galametz2014, Gordon2014, Paradis2019}.
Experimentally, varying far-IR slopes have been observed in laboratory work\cite{Demyk2012, Demyk2022} and are likely due to different temperatures and compositions of dust grains.
This problem has been investigated before, namely using modified blackbody modeling, where the dust opacity (extinction cross-section per unit mass) is in the form $\kappa_{\lambda} \propto \left ( \frac{\lambda}{\lambda_0} \right )^{-\beta}$.
In that case, the far-IR slope is determined by the dust grain spectral index, $\beta$. 
In M31, in particular, Smith et al. (2012, [\citenum{Smith2012}]) provide resolved maps of the dust temperature, dust mass surface density, and spectral index at $36''$ resolution, using far-IR data. Their work shows radial trends of the $\beta$ parameter, increasing from the center to a radius of $\sim 3~$kpc and decreasing to the outskirts. A similar trend was observed by Whitworth et al. (2019, [\citenum{Whitworth2019}]) in the same galaxy with a different modeling framework, and by Tabatabaei et al. (2014, [\citenum{Tabatabaei2014}]) in M33.
These galaxies and the Small and Large Magellanic Clouds (SMC, LMC) were included in the work of Clark et al. (2023, [\citenum{Clark2023}]), using ``two far-IR slopes'' modified blackbody models (so called broken-emissivity models). Radial trends become less pronounced especially in the lower-metallicity systems.
Assuming that temperature variations (correlated with $\beta$) should be limited to the diffuse ISM, we argue that the $\beta$ variation could be predominantly driven by a varying silicate-to-carbon ratio, and investigate how to recover this parameter.
In Chastenet et al. (2017, [\citenum{Chastenet2017}]), the silicate-to-carbon ratio of large grains is tentatively inferred in the SMC and LMC using a physical model on the infrared Spitzer and Herschel SED. They find that the abundance ratio can vary significantly between galaxies, and show spatial variations, hinting at the impact of both the global environment, like metallicity, and the local environment, like the radiation field. However, their ratio maps are subject to significant uncertainties, especially in the SMC, due to spectral sampling and model degeneracies.

The issues faced by Chastenet et al. are partly related to the spectral coverage of available data that did not allow to simultaneously constrain the radiation field and dust abundance parameters with good accuracy.
From a modeling point of view, a model characterizing two grain populations from properties that are observationally accessible is critical. 
This is achievable using The Heterogeneous dust Evolution Model for Interstellar Solids\cite{THEMIS} (\themis).
This model consists of a mixture of amorphous silicates and hydrocarbon grains, both with aromatic-rich mantles. 
In this work, we use the diffuse ISM version of \themis incorporated in \dustem\cite{Compiegne2011} (\url{https://www.ias.u-psud.fr/DUSTEM/}). 
In this framework, \themis is described in practice with four grain populations: pyroxene- and olivine-rich grains, both having the same size distribution parameters, large carbon grains, and small grains tracking the aromatic content coating the big grains. The silicate properties are discussed in Kohler et al. (2014, [\citenum{Kohler2014}]), and the hydrocarbon properties in Jones et al. (2012, [\citenum{JonesOpteca}, \citenum{JonesOptecb}, \citenum{JonesOptecc}]).
The grain intrinsic properties are heavily based on laboratory data. Other parameters such as size distributions and mantle thickness are adjusted to best reproduce the Milky Way observations\cite{Jones2013, THEMIS, Ysard2015}.
A more recent version of this model, \themis{\small 2}, presented in Ysard et al. (2024, [\citenum{Ysard2024}]), revisits some critical aspects of the silicate-rich material, investigating the associated infrared emission associated with optical properties of mixtures studied in laboratory setups, and includes polarization information.

\noindent From an observational point of view, using broad-band photometry up to Herschel/SPIRE~500~$\mu$m is not sufficient to disentangle the far-IR emission of large silicate-rich and large carbon-rich grains\cite{Chastenet2017}. Sub-millimeter-to-millimeter data are needed to partially break down the degeneracy. Although Planck offered many (sub-)millimeter photometric bands, its spatial resolution made this question moot even in the nearby Universe. However, the SCUBA-2 instrument on the James Clerk Maxwell Telescope (JCMT) and its two bands at 450~$\mu$m ($\sim 8''$ FWHM) and 850~$\mu$m ($\sim 13''$ FWHM) are a good complement to Herschel data.
The polarizing nature of dust grains, seldom used in dust emission modeling\cite{Chastenet2022}, can also bring useful constraints. The fraction of polarized light depends on the grain sizes, shapes, and composition\cite{OHM1992, DraineFraisse2009}.
In most cases, infrared dust emission modeling is limited by some degree of parameter degeneracies. Here, we focus on those affecting the far-IR slope specifically. 

This paper is thought around the promise of the PRIMA mission (Glenn et al., this issue). Designed to be a cooled, 1.8~m, far-IR space probe, PRIMA will cover similar wavelengths as Herschel in broad-band photometry filters, but is meant to have better sensitivity, breaking the confusion limit\cite{Bethermin2024}. 
On-board, two instruments, FIRESS and PRIMAger, will allow high-resolution spectroscopy and broad-band photometry. 
In particular, PRIMAger will combine spectroscopy-like and polarimetry capabilities. 
The Hyperspectral Imager will allow $R\sim10$ spectral-like sampling between 24 and 84~$\mu$m, critical in this work. 
The far-IR polarimetric measurements of PRIMAger\cite{Dowell2024} will open a whole new dimension of high-resolution, high-sensitivity dust polarized emission observations in the local universe. It consists of four filters centered on 91, 125, 165, 232~$\mu$m with beam size ranging from 9 to $24''$. All of them will be capable of measuring total and polarized intensities, a most promising feature for dust studies.
Although PRIMA will not sample the infrared regime longwards than Herschel/SPIRE~500~$\mu$m, it will bring great complementary data that will yield a different kind of constrain than its predecessor. As mentioned in the text, we rely on the new JCMT/SCUBA2 high-spatial resolution data to sample the sub-mm wavelength.

This paper presents some of the limitations faced at this stage in dust emission modeling, focused on the possible variations of the SED far-IR slope and the difficulties encountered to recover a silicate-to-carbon ratio.
In Section~\ref{sect:modeltest}, we present tests of modeling degeneracies using synthetic data, 
and in particular, how these limit the estimation of a silicate-to-carbon ratio in a typical dust emission modeling scenario.
In Section~\ref{sect:prima}, we focus on the two modes of the PRIMAger instrument (Ciesla et al., this issue) and investigate how they can be used to mitigate the issues presented before.
Finally, in Section~\ref{sect:m31}, we make an observational case for M31 as a target to use for an empirical test of our assumptions and suggestions. We lay down exposure times for PRIMA's instruments needed to yield conclusive results.

\section{Modeling and testing}
\label{sect:modeltest}
To quantify the degeneracies associated to fitting simultaneously two large grain populations, we rely on the addition of the 850~$\mu$m band from SCUBA-2 in a mid- to far-IR SED. 
We use synthetic SEDs using a known set of parameters, which we then fit using a larger set of emission SEDs, from the same model, and investigate the recovery of radiation field and dust species ratio.
The SEDs are created by integrating infrared spectra in several photometric bands using their transmission curves, mimicking observation from space telescopes:
WISE~3.4, WISE~4.6, WISE~12, WISE~22, Herschel/PACS~70, PACS~100, PACS~160, SPIRE~250, and SPIRE~350, and finally SCUBA-2~850, sampling the spectrum at 3.6, 4.6, 12, 22, 70, 100, 160, 250, 350, and 850~$\mu$m,\footnote{Here, the convention is to use the central wavelength following the instrument name, i.e., PACS~70 means the PACS filter that is centered on 70~$\mu$m.} respectively. The Herschel/SPIRE~500 band is omitted because of its resolution, larger than that of SCUBA-2~850.

\begin{figure}
    \centering
    \includegraphics[width=0.75\linewidth]{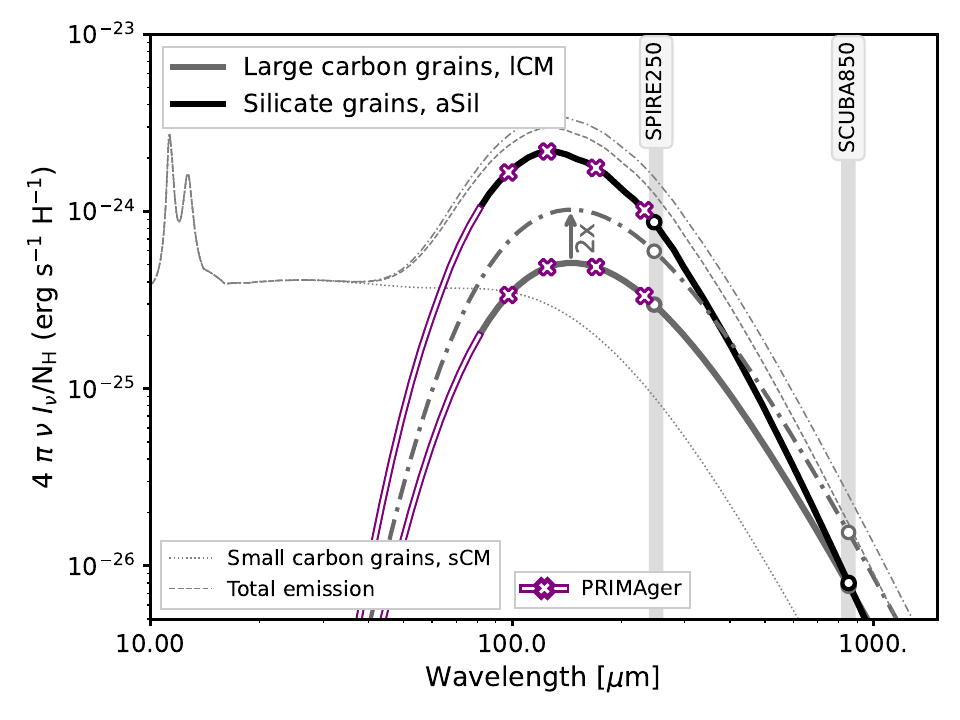}
    \caption[Dust grain emission models from \themis.]{Dust grain emission models from \themis used in this paper, focused on silicate-rich, in black, and carbon-rich, in dark-gray, large grain SEDs. The black and dark-gray solid lines show the emission for the default \themis parametrization. We also show a spectrum for twice the abundance of lCM, in dot-dash line: a break becomes visible at $\sim 400$~$\mu$m, changing the far-IR slope. A change in dust composition is potentially trackable using this part of the emission spectrum.}
    \label{fig:emission}
\end{figure}

\subsection{Parameter definition and mock SEDs}
Using \dustem, we parameterize \themis dust emission SEDs as follow:
\begin{itemize}
\setlength\itemsep{0.02cm}
    \item \umin, the intensity of the radiation field heating the dust grains. \umin scales the local-neighborhood stellar spectrum defined by Mathis et al. (1989, [\citenum{Mathis1983}]) at 10~kpc; 
    \item $\gamma$, the fraction of dust heated by a power-law-integrated range of radiation fields defined as ${\rm U_{min} < U \leq U_{max}}$.
    This parameter is needed to consider the distribution of radiation fields in which dust grains are embedded, the combined effects of multiple stellar sources at different distances from the grains, as well as confusion along the line of sight and resolution effects. The minimum radiation field, \umin, and $\gamma$ are linked by the relation:
    \begin{equation}
        \frac{1}{M_{\rm d}} \left ( \frac{{\rm d}M_{\rm d}}{{\rm d}U} \right ) = 
        (1-\gamma)\ \delta (U - U_{\rm min})\ + 
        \gamma \frac{\alpha - 1}{U_{\rm min}^{1-\alpha} -{\rm U_{max}^{1-\alpha}}}U^{-\alpha},
    \label{EquDustHeating}
    \end{equation}
    where we fix $\alpha=2$ and ${\rm U_{max} = 10^7}$ in this study. The $\delta$ term represents the Dirac function and tracks the $(1-\gamma)$ fraction of total dust mass, $M_{\rm d}$, heated by a single radiation field \umin;
    \item \yasil, the abundance of silicate-rich grains, in which we combine both pyroxene- and olivine-rich materials;
    \item \ylcm, the abundance of large hydrocarbon grains;
    \item \yscm, the abundance of small hydrocarbon grains; `small' and `large' grains' size distributions share the same upper bound and differ in the lower-bounds (grains with 0.4 and 0.5~nm radius, respectively), and different distribution functions: exponential tail and log-normal, respectively. Effectively, large grains become significantly more dominant past $\sim 100~$nm in radius.
    \item \omegastar, the stellar surface brightness that scales a 5\,000~K blackbody to account for stellar emission in the shortest wavelengths (WISE~3.4 and WISE~4.6).
\end{itemize}

In the following, the silicate-to-carbon ratio is defined as $R_{\rm aSil/aC} = Y_{\rm aSil}/Y_{\rm lCM}$. Bear in mind that with this definition, unity is equivalent to the Milky~Way ratio defined by \themis (set to be $(M_{\rm dust}^{\rm aSilM5}/M_{\rm H}) / (M_{\rm dust}^{\rm lCM20}/M_{\rm H}) = 1/0.24 = 4.1$) and not equal mass between silicate-rich and carbon-rich grains. Note that we use the ``diffuse'' version of THEMIS, i.e., calibrated to reproduce the diffuse dust emission at high Galactic latitudes, with no evolution between gas phases, at this point.
We choose not to include the small carbonaceous component in the discussion of the far-IR slope and \rsc. 
Although \yscm bears degeneracies with other parameters, we focus on those affecting the larger grain populations. Under the `diffuse ISM' model, it is reasonable to assume the relative contribution of small grains will not affect the slope as much as larger grains, as shown in Figure~\ref{fig:emission} (see Galliano et al., in this issue, for a study focused on small grain abundance measured with PRIMA).

Figure~\ref{fig:emission} shows a representative SED of \themis, with its default parameters, as set in \dustem. We highlight the emission spectrum of silicate grains (both pyroxene and olivine) in black, and that of the large carbon-rich grains in dark gray. The thick gray dash-dot line shows that same lCM emission spectrum increased by a factor of two, and the associated total spectrum in thinner line. In this figure, we see that a potential break in the far-IR slope happens between the SPIRE~250 and SCUBA-2~850 bands. We therefore use the ratio of these two bands to investigate the far-IR slope variations.
Here we investigate the possibility of using physical dust models to differentiate between silicate and carbon grain emission at long wavelengths.
To understand the results on \rsc, we perform a few tests using synthetic data. 
Based on our parametrization of \themis, we may expect \umin and \rsc to be correlated. This is effectively similar to the degeneracy between temperature and spectral index investigated in previous work\cite{Kelly2012, Galliano2021}. It is a key point of this paper, and an issue PRIMA could resolve. 

\subsection{Variation of the far-IR slope}
\label{sect:modeltest_model}
\begin{figure}
    \centering
    \includegraphics[width=\linewidth, clip, trim={3.8cm 0 3.75cm 0}]{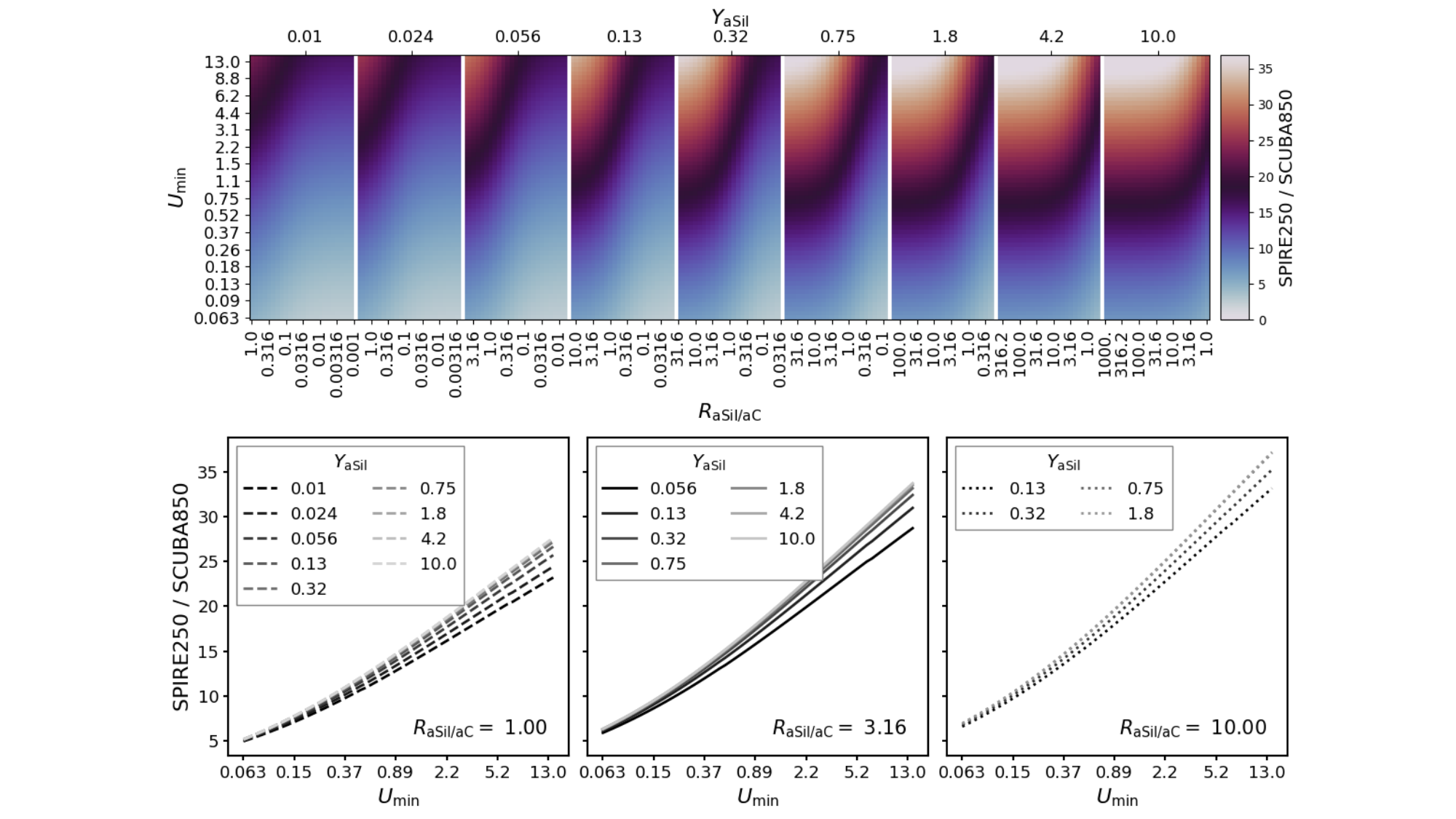}
    \caption[Variation of the far-IR slope.]{Top: Variation of the SPIRE~250/SCUBA-2~850 ratio in \themis. In this image, \yscm, $\gamma$, and \omegastar are fixed. The variations are measured for a range of radiation fields \umin, marked on the y-axis, and several values of \rsc. Some values of the grain ratio are repeated, but are associated with different values of \yasil (split by the vertical white lines), leading to different fluxes, and different flux ratios. The gray lines mark the pixels used in the profiles in the bottom panels.
    Bottom: Profiles of the SPIRE~250/SCUBA-2~850 ratio as a function of \umin, for three different values of \rsc. In each panel, each line is associated with a different \yasil despite sharing the same \rsc. The pixels used for these profiles are marked in the top image by gray lines.}
    \label{fig:slopevar}
\end{figure}

Here, we focus a bit more on the problems linked to recovering the silicate-grain and carbonaceous-grain abundances with our current instrumentation and modeling techniques.
The slope variations are not conspicuous when looking at an emission spectrum, and picturing how all parameters simultaneously affect different parts of the SED is not trivial.
In the top part of Figure~\ref{fig:slopevar}, we show the variations in the ratio of the SPIRE~250-to-SCUBA-2~850 bands, used as a proxy of the far-IR slope, using \themis. 
In the figure, we measure the SPIRE~250/SCUBA-2~850 ratio for a range of \umin, and ranges of amorphous silicate and large carbonaceous grains abundances, while the abundance of small grains, the $\gamma$ parameter, and the stellar surface brightness are fixed. 
On the bottom axis, the \rsc values are sometimes repeated, but are associated with different \yasil values, shown on the top axis, divided in blocks by white vertical lines.

This figure illustrates that the relative flux variation between 250 and 850~$\mu$m depends on more than just the relative abundance of large silicate and carbonaceous grains. For example, the radiation field parameter also has a clear impact, particularly at high \yasil values. This is likely because the impact of increasing \umin varies with the grain material. The heat capacities of silicate-rich or carbon-rich grains are different, and although, in both cases, a higher temperature will shift the infrared emission peak to higher fluxes and shorter wavelengths, the response will not be identical. This may be the reason we observe a change in the far-IR slope with \umin. 
The absolute abundance of grains also seems to matter. We show a better representation of this in the bottom panels of Figure~\ref{fig:slopevar}.
For identical values of \rsc, marked in the top panels by dashed/solid/dotted gray lines, the color gradient in a single column, i.e., as a function of \umin, varies by $\sim 20\%$ for different \yasil values.

From a modeling point of view, this means that recovering the far-IR slope is not only linked to the dust content in the observed pixel. Conclusions drawn from Figure~\ref{fig:slopevar} indicate that properly inferring a silicate-to-carbon ratio implies a correct fit of other parameters as well, namely those related to the radiation field, \umin and, by extension, $\gamma$. 
We explore how PRIMA helps to constrain better the radiation field in Section~\ref{sect:prima_hyperspectral}.

\subsection{Results of the synthetic fit}
\label{sect:modeltest_fit}
For a more quantitative estimation of the uncertainty in recovering a silicate-to-carbon ratio with our current framework, we perform a fit to synthetic data, using common infrared emission modeling We use the previously mentioned wavelengths from WISE and Herschel, at 3.6, 4.6, 12, 22, 70, 100, 160, 250, 350, and 850~$\mu$m.
We first create a few SEDs with known parameters. We create $M\times N$ SEDs for a range of $M$ \umin, and $N$ \rsc, for several values of \yasil, \ylcm (the same way as was done in the previous section).
The other parameters, \yscm, $\gamma$, and \omegastar, are fixed to a single value.
We then create a large library of models to fit these $M\times N$ synthetic SEDs using {\sc dustbff}\cite{Gordon2014, Chastenet2021}. 
{\sc dustbff} measures the likelihood distribution of all the models in the grid for each mock SED provided. It uses covariance matrices to propagate correlated noise between the observing bands, combining instrumental uncertainties (see Clark et al. 2018, [\citenum{Clark2018}], for a useful summary) and background noise. For the latter, we use images of M31 (see Section~\ref{sect:m31}), to measure background noise that fills the diagonal elements of the matrix. We set the off-diagonal elements to 0 for consistency with further tests in the next Section. The best-fit parameter values are measured in the full nD likelihood distribution.
All six parameters are left free in the fitting models.
We make sure that the parameter values in the fitting models do not exactly match those in the synthetic SEDs but are finely sampled, mimicking the unknown truth of astronomical observations.
Some of the values of \rsc are repeated several times, but the associated values of \yasil and \ylcm are different, also leading to different fluxes. 

\begin{figure}
    \centering
    \includegraphics[width=\linewidth]{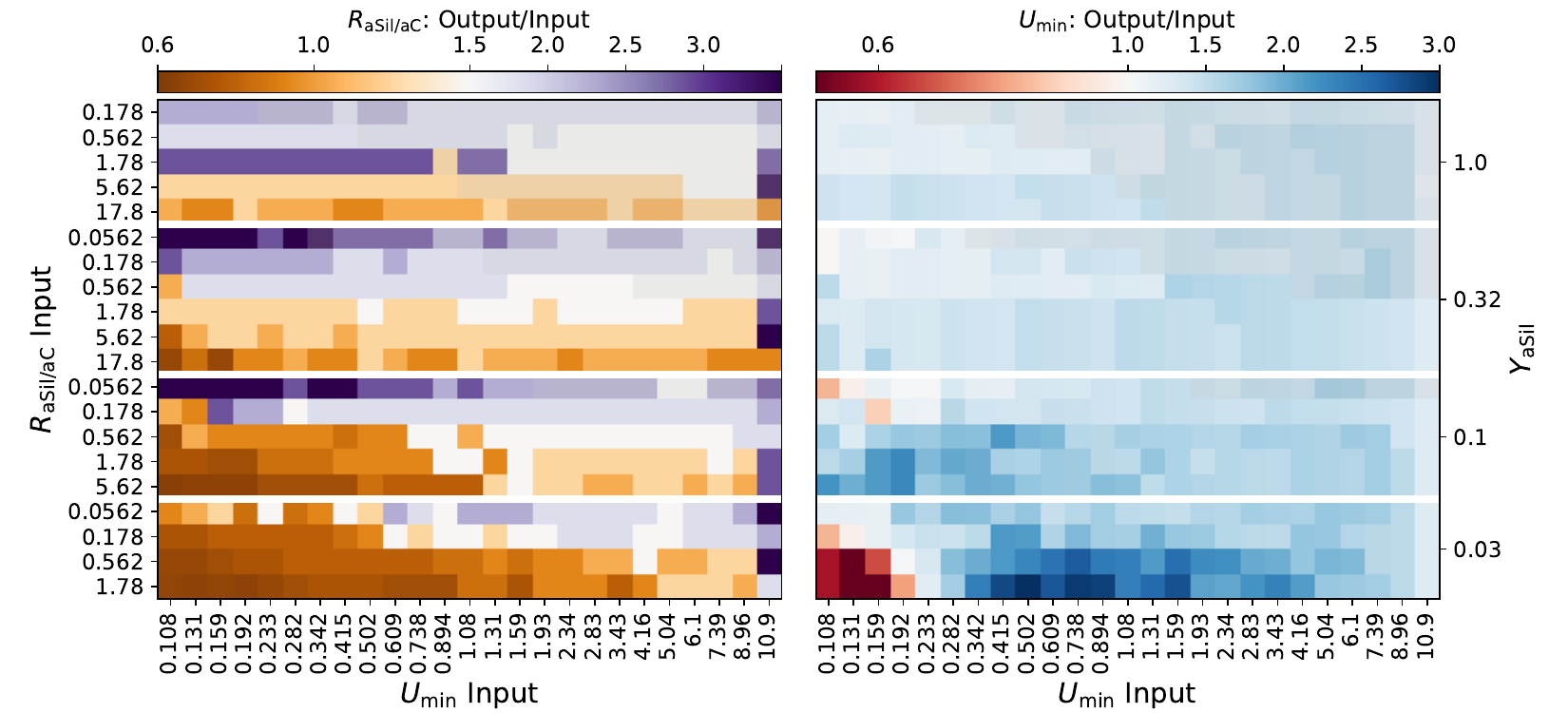}
    \caption[Output-to-input recovery ratios for \rsc and \umin.]{``Recovery'' ratios of \rsc (left panel) and \umin (right panel), after fitting synthetic SEDs with a large grid of \themis emission models. In the left y-axis, the \rsc values are shown, with some of them repeated, but associated with different \yasil values, reported on the opposite axis. We mark these block of unique \yasil values by horizontal white lines.}
    \label{fig:recoveryratios}
\end{figure}

In Figure~\ref{fig:recoveryratios}, we construct the image of the output/input of \rsc and \umin, for all values of \rsc and \umin, reflected on the axes. Each panel is horizontally split in blocks marked by thick white lines, each block sharing a unique \yasil value, reflected on the right-hand-side vertical axis.
It appears more clearly that only a few combinations of (\umin, \rsc) can be properly recovered with this approach.
Note that gray pixels are combinations that create fluxes higher than the maximum fluxes seen in SPIRE~250 and SCUBA-2~850 bands in M31, the observational target upon which we based this study.

In the left panel, we show the ratio of the best-fit value of \rsc (``Out'') to the one in the synthetic SED (``In''). 
In general, it appears that the fit tends to overestimate \rsc, as shown by more orange pixels.
We find that recovering \rsc does not depend only on \rsc itself, but also on the associated values of \yasil and \ylcm (and to a lesser degree on \umin), as indicated with the shifting colors as one moves up the left panel (i.e., higher values of \yasil). 
This is an indication that the quality of the recovery also depends on the surface brightness, rather than only on the slope itself (\rsc), with this set of parameters. This is likely related to the nature of physical dust models, which are often calibrated on high-latitude Milky~Way measurements, and therefore mostly applicable in similar environments, i.e., a Milky~Way-like diffuse ISM.
For each group of a unique \yasil value, there is a clear distinction marked by the contrast of orange and purple colors. This threshold changes with the associated \yasil value, and appears to vary slightly with \umin as well. Despite the default value set by the model, that threshold in \rsc does not seem to be unity, as one might expect, which may indicate that there is a range of (\rsc, \yasil, \umin) that this parametrization can handle and recover. 
We also see that for values of \umin between 0.89 and 8.96, \rsc is globally recovered within $\pm 30$\%.

On the other hand, we do not see this contrast in the right panel, which shows the ratio of the best-fit \umin value to the input value.
In this case, only pixels with the lowest value of \yasil show a particularly poor recovery of the input \umin. 
Some of that behavior is again likely due to the nature of dust models, where \umin is expected to be similar, in that case, to diffuse ISM conditions.
Typically, it is not entirely surprising to see a poor recovery of \umin when it becomes really low, outside of a plausible range of the conditions for which the model is calibrated.

The degeneracies presented in this section make the question of a potentially varying silicate-to-carbon ratio difficult to investigate. This variation is a plausible assumption, based on past work and the different metallicity, stellar population, etc. among the galaxy populations. However, in the current state of dust emission modeling, it is virtually impossible to set up the models without a significant amount of correlation between the model parameters, limiting the conclusions to be drawn.
In the next section, we look at how the upcoming PRIMA instruments capabilities can help overcome this degeneracy issue by bringing more constraints in the two critical regimes we just investigated: radiation field and grain abundances.

\section{PRIMA: helping to solve dust modeling degeneracies}
\label{sect:prima}
The instruments planned for the future PRIMA spacecraft would be of great interest for this particular problem. With greater sensitivity, complementary spectral coverage, and far-IR polarimetry, PRIMA will help us minimize the degeneracies inherent to infrared dust modeling. We build upon the results from Section~\ref{sect:modeltest} to test how PRIMA's instrument can help alleviate the degeneracies in dust modeling previously mentioned.

\subsection{PRIMAger Hyperspectral Imager}
\label{sect:prima_hyperspectral}
\begin{figure}
    \centering
    \includegraphics[width=0.75\linewidth]{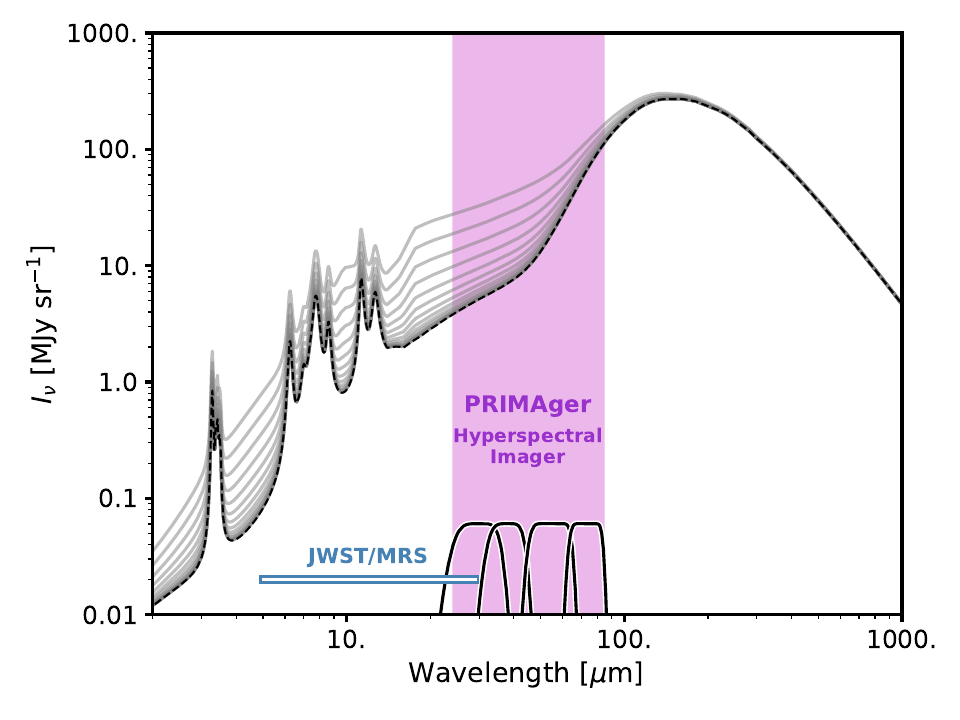}
    \caption[Effects of the $\gamma$ parameter.]{Dust emission SEDs from \themis, with a varying $\gamma$. The black dashed spectrum shows the emission for $U_{\rm min} = 1.0$ and $\gamma=0$. The gray spectra are for the same \umin value and increasing $\gamma$ from $10^{-4}$ to $10^{-1}$. The purple area marks the region probed by PRIMA Hyperspectral imager, which will be critical to help constrain $\gamma$, and by extension \umin. \addtext{The JWST MRS instrument covers $\sim 5$--28~$\mu$m, a synergy that can bring additional constrain to that critical regime.}
    We overlay the four transmission curves we created for the test adding PRIMAger Hyperspectral instrument data points.}
    \label{fig:gamma}
\end{figure}

The Hyperspectral imager onboard PRIMA covering the 24--84~$\mu$m will fill a range of wavelengths for which we lack good spectral sampling. For example, the Spitzer/MIPS instrument had only two bands at 24 and 70~$\mu$m, making it difficult to properly recover the emission in this range. This part of the spectrum is particularly important for estimating the radiation field through the $\gamma$ parameter (Figure~\ref{fig:gamma}), often included in dust emission fitting\cite{Draine2007, Chastenet2025}. It is directly related to the minimum radiation field, \umin, discussed in the previous section, and it controls for the fraction of dust mass that is being heated with radiation fields $U<U_{\rm min}$. The spectral information provided by PRIMA in this range will bring about stronger constraints on the radiation field using information that does not rely on the far-IR slope.

We perform a test to measure the effect of adding information in the 24--84~$\mu$m range. To do so, we use the same set-up outlined in Section~\ref{sect:modeltest_fit}, adding constraints from PRIMA in the mid-IR.
Because of the current {\sc dustbff} set-up (which does not easily allow consideration of a spectrum in the fit) and to keep things straightforward, we do not consider a full 24--84~$\mu$m spectrum but add four data points to our synthetic SEDs in that range. We create mock transmission curves using up-to-date information on the Hyperspectral Imager centered on four wavelengths at $\sim 30$, 38, 55, and 73~$\mu$m, shown in Figure~\ref{fig:gamma}. For the background covariance matrix, we simply interpolate the diagonal elements to the added bands. For the instrument errors, we use a conservative 5\% repeatability error and 10\% correlated error between bands (similar to Herschel errors).

We find that adding four points improves the recovery of both $\gamma$ and \umin (measured as $X_{\rm Out} / X_{\rm In}$). 
We find that \umin is better recovered by $\sim 40\%$, with a scatter decreased by $\sim 27\%$, and $\gamma$ by $\sim 35\%$ with a scatter decreased by $30\%$. 
With this set-up, the mock S/N in the PHI1 band ranges from a few-$\sigma$ to ${\rm S/N} \sim 70$. We find that the residuals in recovering the $\gamma$ parameter decrease with increasing S/N, while \umin seems fairly insensitive to the mock S/N.
We can naturally expect that a full spectrum will yield result even more significant.
However, we also found that the \rsc is more poorly recovered, even though the radiation field parameters are better constrained. 
There can be a few reasons for this, at this stage.
We notice from Figure~\ref{fig:gamma} that $\gamma$ also affects the mid-IR shortwards of 20~$\mu$m, and that may propagate uncertainties in the fit of the small carbon grains population, and eventually affect \rsc. 
We also point out that the choices made to use {\sc dustbff} in a consistent way between tests are bound to affect the recovery of the parameters. Assuming no correlation between all bands possibly hinders the variations in different parts of the spectrum; typically, the WISE mid-IR bands will be correlated with each other without affecting, e.g., the SCUBA2 measurements. For consistency, we used 0s in off-diagonal elements but the empirical truth will differ.
Lastly but likely not least, by adding four points to the 24--84~$\mu$m range, we naturally forced the fit to better recover the parameters that impact this part of the SED. As \yasil and \ylcm will mostly affect the far-IR, the fit tends to favor a model where these two parameters are not as well recovered.
A better handle on spectral input would require to update {\sc dustbff} and/or using other fitting codes that handle spectral input (e.g., Prospector\cite{Prospector}), which is beyond the scope of this paper.

We also believe that a synergy with JWST mid-IR broadband photometry will provide better constraints on the aromatic grain population, responsible for the mid-IR emission features\cite{Sutter2024}. Although we chose not to directly include the smallest grains in the discussion on the silicate-to-carbon ratio, in \themis, they do slightly affect the far-IR emission, and the estimation of \umin, \omegastar, and the mass of grains emitting the mid-IR features are all degenerate.
Using the incredible power of JWST and PRIMA together, it will be possible to properly estimate the abundance of small grains with little degeneracy with the radiation field.

\subsection{PRIMAger Polarimetric Imager}
\label{sect:prima_pola}
The other main aspect for which PRIMA will be tremendously useful is bringing back a far-IR polarimeter. In the past decades, a large amount of work has been done to better understand the theoretical aspect of grain polarization in the ISM of galaxies.
While SOFIA contributed to this progress, its sensitivity and resolution limited the applicability of new theories developed mostly for Milky~Way and very local, bright systems. A broader sample of far-IR polarized emission in now required to further test recent theoretical advances.
The PRIMAger instrument will perfectly sample the far-IR peak in both total and polarized intensity (Figure~\ref{fig:emission}).
The polarized signal depends on the sizes and shapes of aligned dust grains; the new data brought forth by PRIMAger are exactly the ones we need to improve the constraints on individual grain population abundances. At this time, we can expect two main scenarios to analyze the polarization data from PRIMA. 

In the first case, the most simplifying assumption is that carbon grains do not align with the magnetic field and therefore do not contribute to the measured polarized emission. This assumption is somewhat supported by the (possible) lack of ferromagnetic material in purely carbonaceous grains, fully made of carbon and hydrogen atoms. This approach was used by Chastenet et al. (2022, [\citenum{Chastenet2022}]) who use SOFIA/HAWC+ polarized data to derive the dust mass in the Crab supernova remnant.
The power of this assumption is that it greatly simplifies several equations that lead to estimating the dust content traced by the infrared SED. 
However, it is debated in the literature. For example, \themis{\small 2}\cite{Ysard2024} assumes that both silicate and carbon-rich grains align with the magnetic field. This comes from theoretical work by Hoang et al. (2023, [\citenum{Hoang2023}]) suggesting that amorphous carbonaceous grains large enough can contribute significantly to the polarization signal. The exact contribution of each grain population is unknown and would require a precise knowledge of the material that makes up both types. 
Although very simplifying, this approach remains promising: it can provide an upper limit on the silicate grain abundance, which can be used to fit the total emission with better priors. Eventually, this would alleviate, at least in part, the degeneracy between silicate and carbon grain emission in the far-IR.

In the second case, a less stringent approach would be to consider both carbonaceous and silicate grains do polarize the incoming signal, as is intended in \themis{\small 2}. 
However, in order to disentangle both species, one would consecutively solve a series of assumptions, eventually leading to expressing a fraction of either species relative to the other. This would be done using several systems of equations. For example, assuming two different temperatures and spectral indices for carbonaceous and silicate grains, in a modified blackbody emission in both cases, one would use the PRIMAger polarimetry data to derive dust masses of each species.
This (better) approach remains speculative in its details: the literature has not yet reached a consensus on the exact nature of the polarized emission sources, i.e., silicate and carbonaceous grains, and how much each grain population contributes to it.\cite{DraineFraisse2009, HensleyDraine2023, Hoang2023}

\begin{figure}
    \centering
    \includegraphics[width=\linewidth]{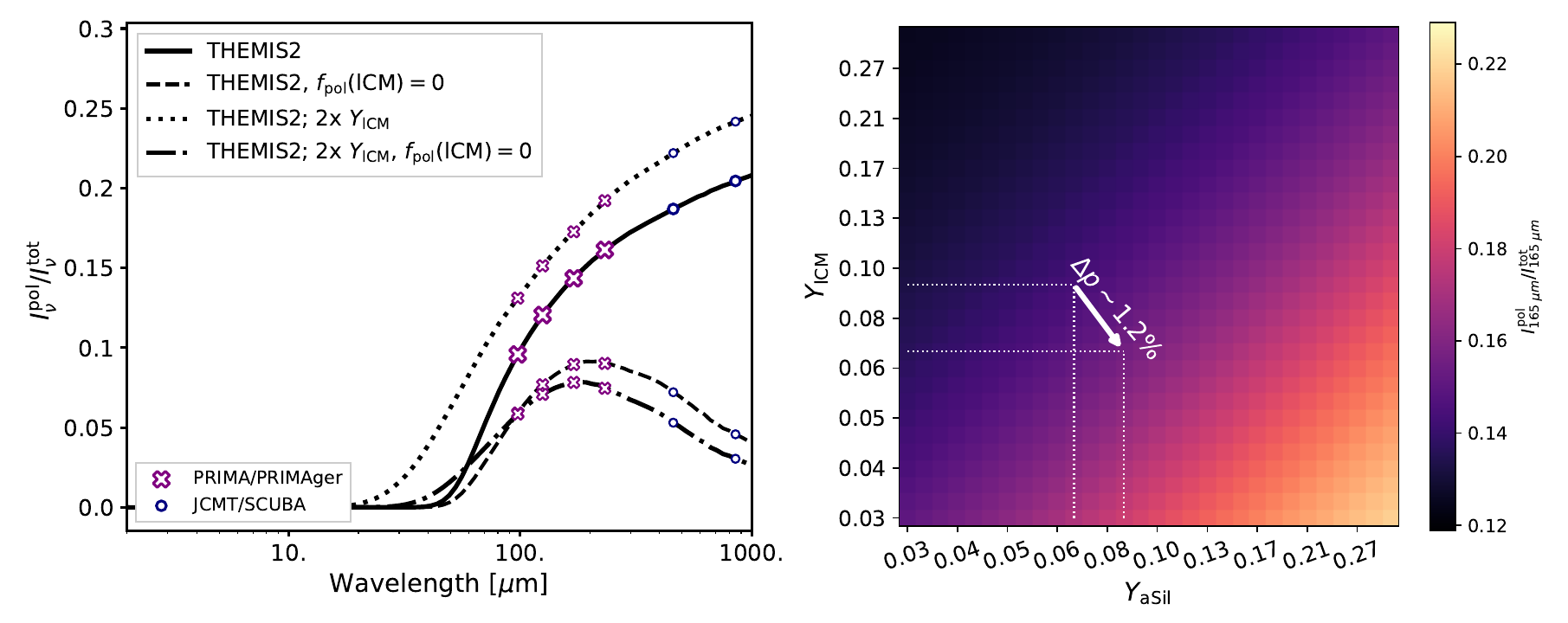}
    \caption[Polarization fraction in \themis{\small 2}.]{\textit{Left:} Polarization fraction as a function of wavelength in \themis{\small 2}. The solid line shows the ``default'' polarization as set in the model, i.e., including polarization from both large carbon and silicate grains; the dash line shows the wavelength-dependent polarization fraction considering only silicate grains polarize light; the dotted line shows the polarization fraction when both silicate and carbon grains contribute, with twice the emission coming from lCM grains; the dash-dotted line also has twice the emission from lCM grains in total intensity, but they do not contribute to the polarized emission. 
    \textit{Right:} Polarization fraction at 165~$\mu$m for the default settings of \themis{\small 2}. We see that measuring a polarization fraction leads to a much restricted range of possible values for \yasil and \ylcm.
    }
    \label{fig:themis2polarisation}
\end{figure}

In Figure~\ref{fig:themis2polarisation}, we show different versions of model predictions from \themis{\small 2} for polarized emission.
In the left panel, we show the polarized fraction from \themis{\small 2}, as a function of wavelength. In the default case, the solid line shows the fraction assuming both grain species align with the magnetic field and polarize light. The dashed line shows the polarized fraction removing all polarized emission contribution from carbonaceous grains. The dotted and dash-dotted lines show the same variations, but for twice the carbon grain abundance.
With the sampling bands from PRIMA's polarimeter, an easy test can be done with future data: the shape of the ``polarization fraction spectrum'' varies significantly depending on the case.
We also add the bands provided by JCMT instrument SCUBA-2 that offers polarization measurements as well.

In the right panel, we build an image of the polarization fraction in the third PRIMAger band. We independently vary the abundances of silicate and carbonaceous grains and measure the polarization fraction at 165~$\mu$m, following the default settings of \themis{\small 2}, i.e., for the solid line in the left panel. In this image, we see that measuring the polarized fraction helps constraining the grain abundances quite significantly. We mark an uncertainty of $\Delta p \sim 1.2\%$ (corresponding to a $3\sigma$ error in our proposed program, in Section~\ref{sect:m31}) and show it strongly limits the range of possible abundances. 
Although that range is well restricted using polarized information, note, however, that the absolute values this constraint leads to is still dependent on the chosen polarized ``model.'' If we applied a different approach, e.g., no polarization from carbonaceous grains, the $Y_{\rm X}$ values would be different. 
This shows that the combination of the several PRIMAger bands will help constrain the shape of the ``polarized spectrum'' and the abundance of grains, which eventually leads to strong information about the silicate-to-carbon ratio.

The new instrumental set-up designed for PRIMA will bring incredible new results in infrared emission modeling. These instruments will fill gaps in spectral regimes and complement existing data by providing polarized emission measurements at high spatial resolution.

\section{The silicate and carbonaceous grain populations of M31}
\label{sect:m31}
In this Section we design an observing program focused on M31. 
For the best outcome of this study, a nearby target is required for better physical resolution, and galaxies in the Local Group are obvious candidates. The Magellanic Clouds, the closest to the Milky~Way, could yield good resolution as well, but they are dwarf, irregular galaxies significantly different from the Milky~Way, in terms of morphology \addtext{and metallicity}. This would add another layer of complexity to interpret the results. Other studies in this issue are focused on the Magellanic Clouds (e.g., Galliano et al.). The other large galaxy, M33, stands a little further and is a little smaller, which simply by argument of distance and size, seems less relevant.
Additionally, using M31 also allows for a more secure distinction between the interarm and the spiral arms environments, compared to M33, where they are a little less clearly defined. M31 also has a higher metallicity, which would lead a better detection in the far-IR bands.
Overall, besides being closer and larger, M31 has also been more studied and has a greater wealth of ancillary data, and we believe it represents a good case for a first solid application of this program\addtext{; at the distance of 765~kpc of M31, the PRIMAger resolution yields $\sim 37$--90~pc in physical scale.}
M31 has been observed by Herschel in its entirety\cite{Fritz2012}, and with the JCMT\cite{Smith2021HASHTAG}, covering the far-IR spectrum and all bands we used in Sections~\ref{sect:modeltest} and \ref{sect:prima}. As mentioned earlier, the spatial properties of the $\beta$ index, tracing the far-IR slope, were studied in several works\cite{Smith2012, Whitworth2019}. 
In the optical, there is the stunning HST imagery of M31 from the PHAT\cite{M31PHATDalcanton2012} and PHAST\cite{M31PHASTChen2025} programs \addtext{(with parsec-scale resolution)}. Stellar population variations inferred from HST could be checked against \rsc spatial variations to link dust production and processing with stellar environment\cite{Gordon2016M31BEAST, Lindberg2024}.
There is also high velocity-resolution H\ {\sc i} mapping of the northern half of M31\cite{M31HIKoch2021}, and CO mapping available as well\cite{M31COCalduPrimo2016}\addtext{, matching the PRIMAger broad-bands FWHM ($18''$, i.e., $\sim 70~$pc, at the VLA resolution)}. These can help separating the ISM into different phases and help inform on the model to be used (e.g., `dense' vs `diffuse' \themis) to track \rsc variations.
As a Milky~Way-like galaxy in an external environment, with an exceptional closeness that allows for highly resolved data, M31 is an ideal laboratory to probe the variations of dust grain properties in extragalactic regions.

\begin{figure}
    \centering
    \includegraphics[width=1.0\linewidth, clip, trim={0 3cm 0 4cm}]{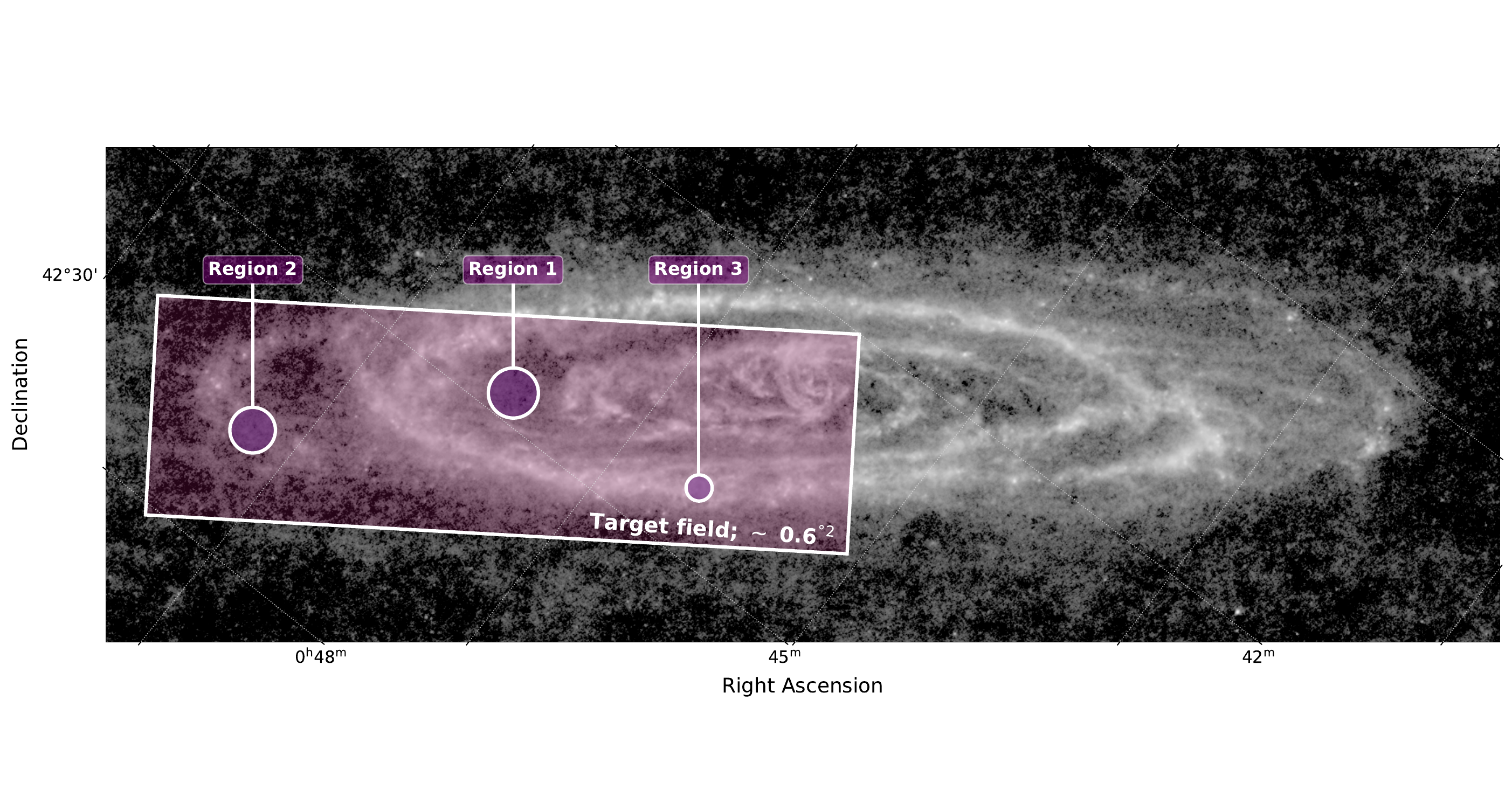}
    \caption[Set-up for the proposed PRIMA observations of M31.]{Herschel/PACS~160 image of M31.
    The rectangle shows the $0.6{{^\circ }^2}$ target region used in the calculation.
    Region~1 is in the interarm and is the reference surface brightness we use for exposure time calculations to yield the desired S/N. Region~2 is a diffuse ``arm,'' visible at 160~$\mu$m but not clearly detected in other bands, and Region~3 is a small bright region in a well-identified arm, for which we provide indicative S/N yielded by our estimated integration times from Region~1.}
    \label{fig:exptime_regs}
\end{figure}

To build upon these works and focus on radial variations and other known properties of dust in M31, we want to sample as many ISM environments as possible. To keep exposure time reasonable, we select a $\sim 0.6\degree$$^2$ area, shown in
Figure~\ref{fig:exptime_regs}, to be observed with PRIMA's instruments, with two main goals:
\begin{itemize}
    \item new radiation field estimates. As seen in  Section~\ref{sect:prima_hyperspectral}, the 24--84~$\mu$m measurements PRIMA would not directly constrain the silicate-to-carbon ratio but will help estimate the radiation field and $\gamma$ parameter. This in turn is expected to alleviate degeneracies in the far-IR.
    In our program, we include the bulge of the galaxy, which is known to have a softer radiation field\cite{Draine2014}. This specific environment will be an excellent region to see the effect of a better radiation field estimate between the older bulge and more diffuse ISM. 
    \item radial variations of \rsc. As discussed in Section~\ref{sect:prima_pola}, the polarimeter will provide high spatial resolution maps of polarized far-IR emission, directly bringing constraints on the dust species abundances.
    The selected area covers radii from the center to the outskirts of the galaxy. This will allow us to compare radial trends of the target \rsc with previous work, namely that of the mentioned spectral index, $\beta$.
\end{itemize} 

In Table~\ref{tab:exptimes}, we present exposure time calculations based on planned characteristics and sensitivities of the PRIMAger instrument, for the highlighted rectangular region.
To do so, we build an average infrared SED from the fit provided by Chastenet et al. (2025, DOI: \href{https://www.ipac.caltech.edu/doi/10.26131/IRSA581}{{\tt 10.26131/IRSA581}}; [\citenum{Chastenet2025}]), in a few pixels in an interarm region, labeled Region~1 shown in Figure~\ref{fig:exptime_regs}. 
From that spectrum, we extract the surface brightness at the four central wavelengths of the polarimetry imager bands, as well as reference values for the hyperspectral imager. 
For total intensity in the polarimetry imager, we calculate the needed exposure time for a target $5\sigma$ detection of the average surface brightness from Region~1, to be measured over the $0.6{{^\circ }^2}$ rectangular region. The same region is used for the hyperspectral imager but we aim for a $3\sigma$ detection due to worse sensitivity.
The exposure times for the polarized emission are calculated with a conservative 2\% polarization fraction. Other work using SOFIA and focused on the spiral, face-on galaxy M51 found polarization fraction between 0.6 and 9\% in the far-IR\cite{Jones2020}. We believe $p\sim2\%$ for M31 is a reasonable target.
The final exposure times amount to \addtext{$\sim 7~$hours} for the Polarimetric Imager and \addtext{$\sim 5$~hours} for the Hyperspectral Imager, aiming for high S/N in M31.

\begin{table}[ht]
\caption[Exposure time calculations for M31.]{Exposure time calculations for the polarimetry imager in total and polarized intensity, assuming 2\% polarization in the latter case. Note that for each, the region used for reference is not the same, due to lower sensitivity in polarimetry mode. For the hyperspectral imager, we used MIPS~24 and PACS~70 maps to assess a reasonable surface brightness to detect at a $3\sigma$ level. 
} 
\label{tab:exptimes}
\begin{center}       
\begin{tabular}{||l|l|l|l|l||} 
\hline\hline
\multicolumn{5}{||c||}{Polarimetry Imager, Total Intensity} \\
\rule[-1ex]{0pt}{3.5ex}  Band & Detection goal & Exposure time, S/N$_{\rm Reg.~1}=5\sigma$ & S/N$_{\rm Reg.~2}$ & S/N$_{\rm Reg.~3}$  \\
\hline
\rule[-1ex]{0pt}{3.5ex}  PPI1 &  12.1 MJy/sr & 1.0~min & 5$\sigma$ & 14$\sigma$ \\
\rule[-1ex]{0pt}{3.5ex}  PPI2 & 13.5 MJy/sr & 0.25~min & 6$\sigma$ & 21$\sigma$ \\
\rule[-1ex]{0pt}{3.5ex}  PPI3 & 10.5 MJy/sr & 0.20~min & 6$\sigma$ & 31$\sigma$ \\
\rule[-1ex]{0pt}{3.5ex}  PPI4 & 6.1 MJy/sr & 0.30~min & 7$\sigma$ & 41$\sigma$ \\
\hline
\multicolumn{5}{||c||}{ } \\
\multicolumn{5}{||c||}{Polarimetry Imager, Polarized Intensity} \\
\rule[-1ex]{0pt}{3.5ex}  Band & Detection goal & Exposure time, S/N$_{\rm Reg.~3}=5\sigma$ & S/N$_{\rm Reg.~1}$ & S/N$_{\rm Reg.~2}$  \\
\hline
\rule[-1ex]{0pt}{3.5ex}  PPI1 &  0.7 MJy/sr & 5.3~hr & 2$\sigma$ & 2$\sigma$ \\
\rule[-1ex]{0pt}{3.5ex}  PPI2 & 1.1 MJy/sr & 1.0~hr & 1$\sigma$ & 1$\sigma$ \\
\rule[-1ex]{0pt}{3.5ex}  PPI3 & 1.3 MJy/sr & 0.44~hr & 1$\sigma$ & 1$\sigma$ \\
\rule[-1ex]{0pt}{3.5ex}  PPI4 & 1.0 MJy/sr & 0.37~hr & 1$\sigma$ & 1$\sigma$ \\
\hline
\hline
\multicolumn{5}{||c||}{ } \\
\multicolumn{5}{||c||}{Hyperspectral Imager} \\
\rule[-1ex]{0pt}{3.5ex}  Band & Detection goal & Exposure time & \multicolumn{2}{c||}{Target S/N}  \\
\hline
\rule[-1ex]{0pt}{3.5ex}  PH1 & 1.25 MJy/sr & 3.7~hr & \multicolumn{2}{l||}{3$\sigma$} \\
\rule[-1ex]{0pt}{3.5ex}  PH2 & 1.25 MJy/sr & 0.75~hr & \multicolumn{2}{l||}{3$\sigma$} \\
\hline\hline
\end{tabular}
\end{center}
\end{table} 

New PRIMA data will grow a set of data already colossal. The ancillary treasure at hand will be complemented by a new kind of data that will greatly advance the empirical tests of dust emission polarization.

\section{Conclusions}
In this paper, we investigate the modeling of the dust emission far-IR slope and how the future PRIMA mission will help leverage some of the current limitations we face. This work aims to measure the variation of the silicate-to-carbon ratio, as traced by the far-IR slope of dust emission. To do so, we use The Heterogeneous dust Evolution Model for Interstellar Solids\cite{THEMIS} (\themis) and update (\themis{\small 2}).

Using physical dust models, we first delve into the parameter degeneracies when considering the ratio SPIRE~250/SCUBA-2~850 photometric bands. This is done allowing two grain populations to vary independently, namely carbonaceous and silicate-rich grains. We find the band ratio naturally varies with the silicate-to-carbon grain abundance ratio, \rsc, but also with the minimum radiation field, \umin. We also find that the absolute abundances of grain species (and not simply their relative abundance) matter significantly (Section~\ref{sect:modeltest_model}, Figure~\ref{fig:slopevar}).

When fitting synthetic data with a common approach, we find that the recovery of known parameters such as the silicate-to-carbon ratio or \umin is good only for certain parameter ranges (Section~\ref{sect:modeltest_fit}, Figure~\ref{fig:recoveryratios}). This is likely due to the nature of dust physical models, which are calibrated to fit measurements of the Milky~Way, and therefore mostly adapted to reproduce similar conditions. However, the main issue remains the degeneracies between the mentioned parameters.

We explore how the instruments on-board PRIMA will be able to help solve this issue, namely its PRIMAger. We find that the critical range covered by PRIMAger's Hyperspectral Imager, from 24 to 84~$\mu$m, will sample a regime that is particularly important to improve the estimate of the radiation field parameter. we find a $> 30\%$ increase in better recovering the \umin and $\gamma$ parameter after adding only four synthetic data points in the Hyperspectral Imager range (Section~\ref{sect:prima_hyperspectral}). 
Similarly, we find that using the capabilities of the Polarimetric Imager can strongly constrain the range of silicate and carbonaceous grain abundances by measuring the far-IR polarization fraction (Section~\ref{sect:prima_pola}, Figure~\ref{fig:themis2polarisation}). Combined with theoretical and laboratory-based work, the observational constraints brought by PRIMA will be tremendously helpful in measuring the variations of \rsc. 

We design an observing proposal using PRIMA's recent specifications. We choose M31 as the ideal target for a first empirical application of our modeling tests. As the largest, closest, spiral galaxy of the Local Group, with a plethora of ancillary data, and past work focused on far-IR slope variations, M31 is the best candidate for our tests. We select a $0.6{\degree}{^2}$ region covering the bulge to the outskirts of the galaxy to probe radial variations of the \rsc, and compute exposure times for a high-S/N set of measurements, as a practical example of the incredible upcoming opportunities PRIMA will offer the interstellar dust astronomy community.

\section*{Disclosure}
The authors declare there are no financial interests, commercial affiliations, or other potential conflicts of interest that have influenced the objectivity of this research or the writing of this paper

\section*{Data Availability}
The data presented in this paper were obtained from open source codes (DustEM, THEMIS, THEMIS2). 

\section*{Acknowledgments}
We thank the referees for their careful reading and suggestions that improved the clarify and flow of this paper.
We thank Karl D. Gordon for useful discussions and advice using {\sc dustbff }.
JC, IDL, and MB acknowledge funding from the Belgian Science Policy Office (BELSPO) through the PRODEX project ``JWST/MIRI Science exploitation'' (C4000142239). IDL acknowledges funding support from the European Research Council (ERC) under the European Union’s Horizon 2020 research and innovation programme DustOrigin (ERC-2019-StG-851622). MR acknowledges support from project PID2023-150178NB-I00 financed by MCIU/AEI/10.13039/501100011033, and by FEDER, UE. 
FG acknowledges support by the French National Research Agency under the contracts WIDENING (ANR-23-ESDIR-0004) and REDEEMING (ANR-24-CE31-2530), as well as by the Actions Thématiques ``Physique et Chimie du Milieu Interstellaire'' (PCMI) of CNRS/INSU, with INC and INP, and ``Cosmologie et Galaxies'' (ATCG) of CNRS/INSU, with INP and IN2P3, both programs being co-funded by CEA and CNES.

\bibliography{report}   

\begin{thebibliography}{10}

\bibitem{DeLooze2014}
I.~{De Looze}, J.~{Fritz}, M.~{Baes}, {\em et~al.}, ``{High-resolution, 3D radiative transfer modeling. I. The grand-design spiral galaxy M 51},'' {\em \aap} {\bf 571}, A69  (2014).

\bibitem{Nersesian2020}
A.~{Nersesian}, S.~{Verstocken}, S.~{Viaene}, {\em et~al.}, ``{High-resolution, 3D radiative transfer modelling. III. The DustPedia barred galaxies},'' {\em \aap} {\bf 637}, A25  (2020).

\bibitem{Nersesian2020M51}
A.~{Nersesian}, S.~{Viaene}, I.~{De Looze}, {\em et~al.}, ``{High-resolution, 3D radiative transfer modelling. V. A detailed model of the M 51 interacting pair},'' {\em \aap} {\bf 643}, A90  (2020).

\bibitem{Verstocken2020}
S.~{Verstocken}, A.~{Nersesian}, M.~{Baes}, {\em et~al.}, ``{High-resolution, 3D radiative transfer modelling. II. The early-type spiral galaxy M 81},'' {\em \aap} {\bf 637}, A24  (2020).

\bibitem{Viaene2020}
S.~{Viaene}, A.~{Nersesian}, J.~{Fritz}, {\em et~al.}, ``{High-resolution, 3D radiative transfer modelling. IV. AGN-powered dust heating in NGC 1068},'' {\em \aap} {\bf 638}, A150  (2020).

\bibitem{Viaene2016}
S.~{Viaene}, M.~{Baes}, G.~{Bendo}, {\em et~al.}, ``{The bolometric and UV attenuation in normal spiral galaxies of the Herschel Reference Survey},'' {\em \aap} {\bf 586}, A13  (2016).

\bibitem{Bianchi2018}
S.~{Bianchi}, P.~{De Vis}, S.~{Viaene}, {\em et~al.}, ``{Fraction of bolometric luminosity absorbed by dust in DustPedia galaxies},'' {\em \aap} {\bf 620}, A112  (2018).

\bibitem{LeBourlot2012}
J.~{Le Bourlot}, F.~{Le Petit}, C.~{Pinto}, {\em et~al.}, ``{Surface chemistry in the interstellar medium. I. H$_{2}$ formation by Langmuir-Hinshelwood and Eley-Rideal mechanisms},'' {\em \aap} {\bf 541}, A76  (2012).

\bibitem{Bron2014}
E.~{Bron}, J.~{Le Bourlot}, and F.~{Le Petit}, ``{Surface chemistry in the interstellar medium. II. H$_{2}$ formation on dust with random temperature fluctuations},'' {\em \aap} {\bf 569}, A100  (2014).

\bibitem{Wolfire95}
M.~G. {Wolfire}, D.~{Hollenbach}, C.~F. {McKee}, {\em et~al.}, ``{The neutral atomic phases of the interstellar medium},'' {\em \apj} {\bf 443}, 152--168  (1995).

\bibitem{Gordon2014}
K.~D. {Gordon}, J.~{Roman-Duval}, C.~{Bot}, {\em et~al.}, ``{Dust and Gas in the Magellanic Clouds from the HERITAGE Herschel Key Project. I. Dust Properties and Insights into the Origin of the Submillimeter Excess Emission},'' {\em \apj} {\bf 797}, 85  (2014).

\bibitem{MeisnerFinkbeiner2015}
A.~M. {Meisner} and D.~P. {Finkbeiner}, ``{Modeling Thermal Dust Emission with Two Components: Application to the Planck High Frequency Instrument Maps},'' {\em \apj} {\bf 798}, 88  (2015).

\bibitem{Zubko2004}
V.~{Zubko}, E.~{Dwek}, and R.~G. {Arendt}, ``{Interstellar Dust Models Consistent with Extinction, Emission, and Abundance Constraints},'' {\em \apjs} {\bf 152}, 211--249  (2004).

\bibitem{DL2007}
B.~T. {Draine} and A.~{Li}, ``{Infrared Emission from Interstellar Dust. IV. The Silicate-Graphite-PAH Model in the Post-Spitzer Era},'' {\em \apj} {\bf 657}, 810--837  (2007).

\bibitem{Jones2013}
A.~P. {Jones}, L.~{Fanciullo}, M.~{K{\"o}hler}, {\em et~al.}, ``{The evolution of amorphous hydrocarbons in the ISM: dust modelling from a new vantage point},'' {\em \aap} {\bf 558}, A62  (2013).

\bibitem{THEMIS}
A.~P. {Jones}, M.~{K{\"o}hler}, N.~{Ysard}, {\em et~al.}, ``{The global dust modelling framework THEMIS},'' {\em \aap} {\bf 602}, A46  (2017).

\bibitem{MunozMateos2009}
J.~C. {Mu{\~n}oz-Mateos}, A.~{Gil de Paz}, S.~{Boissier}, {\em et~al.}, ``{Radial Distribution of Stars, Gas, and Dust in Sings Galaxies. II. Derived Dust Properties},'' {\em \apj} {\bf 701}, 1965--1991  (2009).

\bibitem{Boselli2010}
A.~{Boselli}, S.~{Eales}, L.~{Cortese}, {\em et~al.}, ``{The Herschel Reference Survey},'' {\em \pasp} {\bf 122}, 261  (2010).

\bibitem{Ciesla2014}
L.~{Ciesla}, M.~{Boquien}, A.~{Boselli}, {\em et~al.}, ``{Dust spectral energy distributions of nearby galaxies: an insight from the Herschel Reference Survey},'' {\em \aap} {\bf 565}, A128  (2014).

\bibitem{RemyRuyer2014}
A.~{R{\'e}my-Ruyer}, S.~C. {Madden}, F.~{Galliano}, {\em et~al.}, ``{Gas-to-dust mass ratios in local galaxies over a 2 dex metallicity range},'' {\em \aap} {\bf 563}, A31  (2014).

\bibitem{Hunt2015}
L.~K. {Hunt}, B.~T. {Draine}, S.~{Bianchi}, {\em et~al.}, ``{Cool dust heating and temperature mixing in nearby star-forming galaxies},'' {\em \aap} {\bf 576}, A33  (2015).

\bibitem{Davies2017}
J.~I. {Davies}, M.~{Baes}, S.~{Bianchi}, {\em et~al.}, ``{DustPedia: A Definitive Study of Cosmic Dust in the Local Universe},'' {\em \pasp} {\bf 129}, 044102  (2017).

\bibitem{Aniano2020}
G.~{Aniano}, B.~T. {Draine}, L.~K. {Hunt}, {\em et~al.}, ``{Modeling Dust and Starlight in Galaxies Observed by Spitzer and Herschel: The KINGFISH Sample},'' {\em \apj} {\bf 889}, 150  (2020).

\bibitem{Galliano2021}
F.~{Galliano}, A.~{Nersesian}, S.~{Bianchi}, {\em et~al.}, ``{A nearby galaxy perspective on dust evolution. Scaling relations and constraints on the dust build-up in galaxies with the DustPedia and DGS samples},'' {\em \aap} {\bf 649}, A18  (2021).

\bibitem{Abdurrouf2022}
{Abdurro'uf}, Y.-T. {Lin}, H.~{Hirashita}, {\em et~al.}, ``{Dissecting Nearby Galaxies with piXedfit. II. Spatially Resolved Scaling Relations among Stars, Dust, and Gas},'' {\em \apj} {\bf 935}, 98  (2022).

\bibitem{Casasola2022}
V.~{Casasola}, S.~{Bianchi}, L.~{Magrini}, {\em et~al.}, ``{The resolved scaling relations in DustPedia: Zooming in on the local Universe},'' {\em \aap} {\bf 668}, A130  (2022).

\bibitem{Dale2023}
D.~A. {Dale}, M.~{Boquien}, J.~A. {Turner}, {\em et~al.}, ``{Spectral Energy Distributions for 258 Local Volume Galaxies},'' {\em \aj} {\bf 165}, 260  (2023).

\bibitem{Chastenet2025}
J.~{Chastenet}, K.~{Sandstrom}, A.~K. {Leroy}, {\em et~al.}, ``{The Resolved Behavior of Dust Mass, Polycyclic Aromatic Hydrocarbon Fraction, and Radiation Field in {\ensuremath{\sim}}800 Nearby Galaxies},'' {\em \apjs} {\bf 276}, 2  (2025).

\bibitem{Gail2009}
H.~P. {Gail}, S.~V. {Zhukovska}, P.~{Hoppe}, {\em et~al.}, ``{Stardust from Asymptotic Giant Branch Stars},'' {\em \apj} {\bf 698}, 1136--1154  (2009).

\bibitem{Goldman2022}
S.~R. {Goldman}, M.~L. {Boyer}, J.~{Dalcanton}, {\em et~al.}, ``{A Census of Thermally Pulsing AGB Stars in the Andromeda Galaxy and a First Estimate of Their Contribution to the Global Dust Budget},'' {\em \apjs} {\bf 259}, 41  (2022).

\bibitem{Bocchio2014}
M.~{Bocchio}, A.~P. {Jones}, and J.~D. {Slavin}, ``{A re-evaluation of dust processing in supernova shock waves},'' {\em \aap} {\bf 570}, A32  (2014).

\bibitem{Slavin2015}
J.~D. {Slavin}, E.~{Dwek}, and A.~P. {Jones}, ``{Destruction of Interstellar Dust in Evolving Supernova Remnant Shock Waves},'' {\em \apj} {\bf 803}, 7  (2015).

\bibitem{Hu2019}
C.-Y. {Hu}, S.~{Zhukovska}, R.~S. {Somerville}, {\em et~al.}, ``{Thermal and non-thermal dust sputtering in hydrodynamical simulations of the multiphase interstellar medium},'' {\em \mnras} {\bf 487}, 3252--3269  (2019).

\bibitem{Bot2010}
C.~{Bot}, N.~{Ysard}, D.~{Paradis}, {\em et~al.}, ``{Submillimeter to centimeter excess emission from the Magellanic Clouds. II. On the nature of the excess},'' {\em \aap} {\bf 523}, A20  (2010).

\bibitem{Galliano2011}
F.~{Galliano}, S.~{Hony}, J.~P. {Bernard}, {\em et~al.}, ``{Non-standard grain properties, dark gas reservoir, and extended submillimeter excess, probed by Herschel in the Large Magellanic Cloud},'' {\em \aap} {\bf 536}, A88  (2011).

\bibitem{Galametz2014}
M.~{Galametz}, M.~{Albrecht}, R.~{Kennicutt}, {\em et~al.}, ``{Dissecting the origin of the submillimetre emission in nearby galaxies with Herschel and LABOCA},'' {\em \mnras} {\bf 439}, 2542--2570  (2014).

\bibitem{Paradis2019}
D.~{Paradis}, C.~{M{\'e}ny}, M.~{Juvela}, {\em et~al.}, ``{Revisiting the dust properties in the molecular clouds of the Large Magellanic Cloud},'' {\em \aap} {\bf 627}, A15  (2019).

\bibitem{Demyk2012}
K.~{Demyk}, C.~{Meny}, A.~{Coupeaud}, {\em et~al.}, ``{Variation of the FIR/submm optical properties of silicate dust analogues at low temperature: implications for the observations of interstellar cold dust emission},'' in {\em EAS Publications Series},  C.~{Stehl{\'e}}, C.~{Joblin}, and L.~{d'Hendecourt}, Eds., {\em EAS Publications Series} {\bf 58}, 405--408  (2012).

\bibitem{Demyk2022}
K.~{Demyk}, C.~{Meny}, X.~H. {Lu}, {\em et~al.}, ``{Low temperature MIR to submillimeter mass absorption coefficient of interstellar dust analogues. I. Mg-rich glassy silicates (Corrigendum)},'' {\em \aap} {\bf 666}, C3  (2022).

\bibitem{Smith2012}
M.~W.~L. {Smith}, S.~A. {Eales}, H.~L. {Gomez}, {\em et~al.}, ``{The Herschel Exploitation of Local Galaxy Andromeda (HELGA). II. Dust and Gas in Andromeda},'' {\em \apj} {\bf 756}, 40  (2012).

\bibitem{Whitworth2019}
A.~P. {Whitworth}, K.~A. {Marsh}, P.~J. {Cigan}, {\em et~al.}, ``{The dust in M31},'' {\em \mnras} {\bf 489}, 5436--5452  (2019).

\bibitem{Tabatabaei2014}
F.~S. {Tabatabaei}, J.~{Braine}, E.~M. {Xilouris}, {\em et~al.}, ``{Variation in the dust emissivity index across M 33 with Herschel and Spitzer (HerM 33es)},'' {\em \aap} {\bf 561}, A95  (2014).

\bibitem{Clark2023}
C.~J.~R. {Clark}, J.~C. {Roman-Duval}, K.~D. {Gordon}, {\em et~al.}, ``{The Quest for the Missing Dust. II. Two Orders of Magnitude of Evolution in the Dust-to-gas Ratio Resolved within Local Group Galaxies},'' {\em \apj} {\bf 946}, 42  (2023).

\bibitem{Chastenet2017}
J.~{Chastenet}, C.~{Bot}, K.~D. {Gordon}, {\em et~al.}, ``{Modeling dust emission in the Magellanic Clouds with Spitzer and Herschel},'' {\em \aap} {\bf 601}, A55  (2017).

\bibitem{Compiegne2011}
M.~{Compi{\`e}gne}, L.~{Verstraete}, A.~{Jones}, {\em et~al.}, ``{The global dust SED: tracing the nature and evolution of dust with DustEM},'' {\em \aap} {\bf 525}, A103  (2011).

\bibitem{DustEMurl}
D.~T. Institut~d'Astrophysique Spatiale, ``Dustem, a modelling tool for dust emission and extinction.'' \url{https://www.ias.u-psud.fr/DUSTEM/}  (2013).

\bibitem{Kohler2014}
M.~{K{\"o}hler}, A.~{Jones}, and N.~{Ysard}, ``{A hidden reservoir of Fe/FeS in interstellar silicates?},'' {\em \aap} {\bf 565}, L9  (2014).

\bibitem{JonesOpteca}
A.~P. {Jones}, ``{Variations on a theme - the evolution of hydrocarbon solids. I. Compositional and spectral modelling - the eRCN and DG models},'' {\em \aap} {\bf 540}, A1  (2012).

\bibitem{JonesOptecb}
A.~P. {Jones}, ``{Variations on a theme - the evolution of hydrocarbon solids. II. Optical property modelling - the optEC$_{(s)}$ model},'' {\em \aap} {\bf 540}, A2  (2012).

\bibitem{JonesOptecc}
A.~P. {Jones}, ``{Variations on a theme - the evolution of hydrocarbon solids. III. Size-dependent properties - the optEC$_{(s)}$(a) model},'' {\em \aap} {\bf 542}, A98  (2012).

\bibitem{Ysard2015}
N.~{Ysard}, M.~{K{\"o}hler}, A.~{Jones}, {\em et~al.}, ``{Dust variations in the diffuse interstellar medium: constraints on Milky Way dust from Planck-HFI observations},'' {\em \aap} {\bf 577}, A110  (2015).

\bibitem{Ysard2024}
N.~{Ysard}, A.~P. {Jones}, V.~{Guillet}, {\em et~al.}, ``{THEMIS 2.0: A self-consistent model for dust extinction, emission, and polarisation},'' {\em \aap} {\bf 684}, A34  (2024).

\bibitem{Chastenet2022}
J.~{Chastenet}, I.~{De Looze}, B.~S. {Hensley}, {\em et~al.}, ``{SOFIA/HAWC+ observations of the Crab Nebula: dust properties from polarized emission},'' {\em \mnras} {\bf 516}, 4229--4244  (2022).

\bibitem{OHM1992}
V.~{Ossenkopf}, T.~{Henning}, and J.~S. {Mathis}, ``{Constraints on cosmic silicates.},'' {\em \aap} {\bf 261}, 567--578  (1992).

\bibitem{DraineFraisse2009}
B.~T. {Draine} and A.~A. {Fraisse}, ``{Polarized Far-Infrared and Submillimeter Emission from Interstellar Dust},'' {\em \apj} {\bf 696}, 1--11  (2009).

\bibitem{Bethermin2024}
M.~{B{\'e}thermin}, A.~D. {Bolatto}, F.~{Boulanger}, {\em et~al.}, ``{Confusion of extragalactic sources in the far-infrared: A baseline assessment of the performance of PRIMAger in intensity and polarization},'' {\em \aap} {\bf 692}, A52  (2024).

\bibitem{Dowell2024}
C.~D. {Dowell}, B.~S. {Hensley}, and M.~{Sauvage}, ``{Simulation of the Far-Infrared Polarimetry Approach Envisioned for the PRIMA Mission},'' {\em arXiv e-prints} , arXiv:2404.17050  (2024).

\bibitem{Mathis1983}
J.~S. {Mathis}, P.~G. {Mezger}, and N.~{Panagia}, ``{Interstellar radiation field and dust temperatures in the diffuse interstellar medium and in giant molecular clouds},'' {\em \aap} {\bf 128}, 212--229  (1983).

\bibitem{Kelly2012}
B.~C. {Kelly}, R.~{Shetty}, A.~M. {Stutz}, {\em et~al.}, ``{Dust Spectral Energy Distributions in the Era of Herschel and Planck: A Hierarchical Bayesian-fitting Technique},'' {\em \apj} {\bf 752}, 55  (2012).

\bibitem{Chastenet2021}
J.~{Chastenet}, K.~{Sandstrom}, I.-D. {Chiang}, {\em et~al.}, ``{Benchmarking Dust Emission Models in M101},'' {\em \apj} {\bf 912}, 103  (2021).

\bibitem{Clark2018}
C.~J.~R. {Clark}, S.~{Verstocken}, S.~{Bianchi}, {\em et~al.}, ``{DustPedia: Multiwavelength photometry and imagery of 875 nearby galaxies in 42 ultraviolet-microwave bands},'' {\em \aap} {\bf 609}, A37  (2018).

\bibitem{Draine2007}
B.~T. {Draine}, D.~A. {Dale}, G.~{Bendo}, {\em et~al.}, ``{Dust Masses, PAH Abundances, and Starlight Intensities in the SINGS Galaxy Sample},'' {\em \apj} {\bf 663}, 866--894  (2007).

\bibitem{Prospector}
J.~{Leja}, B.~D. {Johnson}, C.~{Conroy}, {\em et~al.}, ``{Deriving Physical Properties from Broadband Photometry with Prospector: Description of the Model and a Demonstration of its Accuracy Using 129 Galaxies in the Local Universe},'' {\em \apj} {\bf 837}, 170  (2017).

\bibitem{Sutter2024}
J.~{Sutter}, K.~{Sandstrom}, J.~{Chastenet}, {\em et~al.}, ``{The Fraction of Dust Mass in the Form of Polycyclic Aromatic Hydrocarbons on 10{\textendash}50 pc Scales in Nearby Galaxies},'' {\em \apj} {\bf 971}, 178  (2024).

\bibitem{Hoang2023}
T.~{Hoang}, V.~H. {Minh Phan}, and L.~N. {Tram}, ``{Internal and External Alignment of Carbonaceous Grains within the Radiative Torque Paradigm},'' {\em \apj} {\bf 954}, 216  (2023).

\bibitem{HensleyDraine2023}
B.~S. {Hensley} and B.~T. {Draine}, ``{The Astrodust+PAH Model: A Unified Description of the Extinction, Emission, and Polarization from Dust in the Diffuse Interstellar Medium},'' {\em \apj} {\bf 948}, 55  (2023).

\bibitem{Fritz2012}
J.~{Fritz}, G.~{Gentile}, M.~W.~L. {Smith}, {\em et~al.}, ``{The Herschel Exploitation of Local Galaxy Andromeda (HELGA). I. Global far-infrared and sub-mm morphology},'' {\em \aap} {\bf 546}, A34  (2012).

\bibitem{Smith2021HASHTAG}
M.~W.~L. {Smith}, S.~A. {Eales}, T.~G. {Williams}, {\em et~al.}, ``{The HASHTAG Project: The First Submillimeter Images of the Andromeda Galaxy from the Ground},'' {\em \apjs} {\bf 257}, 52  (2021).

\bibitem{M31PHATDalcanton2012}
J.~J. {Dalcanton}, B.~F. {Williams}, D.~{Lang}, {\em et~al.}, ``{The Panchromatic Hubble Andromeda Treasury},'' {\em \apjs} {\bf 200}, 18  (2012).

\bibitem{M31PHASTChen2025}
Z.~{Chen}, B.~{Williams}, D.~{Lang}, {\em et~al.}, ``{PHAST. The Panchromatic Hubble Andromeda Southern Treasury. I. Ultraviolet and Optical Photometry of over 90 Million Stars in M31},'' {\em \apj} {\bf 979}, 35  (2025).

\bibitem{Gordon2016M31BEAST}
K.~D. {Gordon}, M.~{Fouesneau}, H.~{Arab}, {\em et~al.}, ``{The Panchromatic Hubble Andromeda Treasury. XV. The BEAST: Bayesian Extinction and Stellar Tool},'' {\em \apj} {\bf 826}, 104  (2016).

\bibitem{Lindberg2024}
C.~W. {Lindberg}, C.~E. {Murray}, J.~J. {Dalcanton}, {\em et~al.}, ``{Dust around Massive Stars Is Agnostic to Galactic Environment: New Insights from PHAT/BEAST},'' {\em \apj} {\bf 963}, 58  (2024).

\bibitem{M31HIKoch2021}
E.~W. {Koch}, E.~W. {Rosolowsky}, A.~K. {Leroy}, {\em et~al.}, ``{A lack of constraints on the cold opaque H I mass: H I spectra in M31 and M33 prefer multicomponent models over a single cold opaque component},'' {\em \mnras} {\bf 504}, 1801--1824  (2021).

\bibitem{M31COCalduPrimo2016}
A.~{Cald{\'u}-Primo} and A.~{Schruba}, ``{Molecular Gas Velocity Dispersions in the Andromeda Galaxy},'' {\em \aj} {\bf 151}, 34  (2016).

\bibitem{Draine2014}
B.~T. {Draine}, G.~{Aniano}, O.~{Krause}, {\em et~al.}, ``{Andromeda's Dust},'' {\em \apj} {\bf 780}, 172  (2014).

\bibitem{Jones2020}
T.~J. {Jones}, J.-A. {Kim}, C.~D. {Dowell}, {\em et~al.}, ``{HAWC+ Far-infrared Observations of the Magnetic Field Geometry in M51 and NGC 891},'' {\em \aj} {\bf 160}, 167  (2020).

\end{thebibliography}
\bibliographystyle{spiejour}   


\vspace{2ex}\noindent\textbf{J\'er\'emy Chastenet} is a postdoctoral scholar at Ghent University, Belgium. He received his MS and PhD in astrophysics from the University of Strasbourg, France, with a co-supervision at the Space Telescope Science Institute, Baltimore, MD, United-States. His research interests are focused on observational astronomy probing the interstellar medium of nearby galaxies.

\vspace{2ex}\noindent\textbf{Ilse De Looze} is an associate professor at Ghent University, Belgium. She received her MS and PhD in astronomy from Ghent University, Belgium, in 2009 and 2012, respectively. Her main research area revolves around interstellar dust, using of multi-wavelength observations, chemical and dust evolution models, and numerical dust destruction models to constrain the different stages of the dust lifecycle.

\vspace{2ex}\noindent\textbf{Frédéric Galliano} is a staff researcher at CNRS/AIM associated with the Department of Astrophysics of CEA Paris-Saclay.
He received his MS and PhD degrees in astrophysics from the University of Paris XI in 2000 and 2004, respectively. 
He is the author of about 200 journal papers.
He develops the hierarchical Bayesian dust SED code HerBIE.
His current research interests include interstellar dust grains, and infrared and millimeter observations.

\vspace{1ex}
\noindent Biographies and photographs of the other authors are not available.

\listoffigures
\listoftables

\end{spacing}
\end{document}